\documentclass[12pt,preprint]{aastex}
\usepackage[twocolumn]{emulateapj5}
\usepackage{apjfonts}
\usepackage{graphicx}

\slugcomment{To be submitted to the Astrophysical Journal}
\shorttitle{C and N abundances in Algol B}
\shortauthors{Drake}
\begin{document}
\title{From the Heart of The Ghoul: C and N Abundances in the
Corona of Algol B
}
\author{Jeremy J. Drake\altaffilmark{1}}
\affil{$^1$Smithsonian Astrophysical Observatory,
MS-3, \\ 60 Garden Street, \\ Cambridge, MA 02138}

\begin{abstract}

{\it Chandra} Low Energy Transmission Grating Spectrograph
observations of Algol have been used to determine the abundances of C
and N in the secondary star for the first time.  In order to minimise
errors arising from an uncertain coronal differential emission measure
as a function of temperature, the analysis was performed relative to
similar observations of an adopted ``standard'' star HR~1099.  It is
demonstrated  HR~1099 and Algol are coronal twins in many respects
and that their X-ray spectra are very
similar in nearly all details, except for the observed strengths of C
and N lines.  The H-like 2p~$^2$P$_{3/2,1/2}$ $\rightarrow$
1s~$^2$S$_{1/2}$ transitions of C and N in the coronae of Algol and
HR~1099 demonstrate that the surface abundances of Algol~B have been
strongly modified by CN-processing, as demonstrated earlier by Schmitt
\& Ness (2002).  It is found that N is enhanced in Algol~B by a factor
of 3 compared to HR~1099.  No C lines are detected in the Algol
spectrum, indicating a C depletion relative to HR~1099 by a factor of
10 or more.  These C and N abundances indicate that Algol~B must have
lost at least half of its initial mass, and are consistent with
predictions of evolutionary models that include non-conservative mass
transfer and angular momentum loss through magnetic activity.
Based on H-like and He-like transitions in O and Ne, it
is estimated that Algol is slightly metal-poor by 0.2~dex in terms of
the coronal abundances of light elements relative to HR~1099, while
the Fe~XVII 2p$^5$3d~$^1$P$_1$ $\rightarrow$ 2p$^6$~$^1$S$_0$
transition indicates a very similar Fe abundance.  In reviewing
coronal abundance results for active stars in the literature, and
drawing on an earlier {\it Chandra} study of the coronal abundances of
HR~1099, it is concluded that Fe is very likely depleted in the
coronae of both Algol and HR~1099 by 0.5~dex relative to their
photospheric compositions, but that Ne is enhanced by a similar
magnitude.  Light elements such as C, N and O are likely depleted in
both stars by of order 0.3~dex.  The similarities in these large
abundance anomalies in HR~1099 and Algol is notable.  Despite such
compositional fractionation in these coronae, the relative C and N
abundances in HR~1099, determined by comparing observed line strengths
to theoretical C/N line ratios, are consistent with recent solar
values, indicating that differential fractionation between these
elements is not significant and that little or no dredge-up of
material subjected to CN-processing has occurred on the subgiant
component.

\end{abstract}


\section{Introduction}
\label{s:intro}

Algol, whose name derives from Arabic for ``the ghoul'', is the
brightest eclipsing binary system in the sky and the prototype of its
class.  Algol-type binaries provide laboratories for studying
close-binary evolution, Roche lobe overflow and mass transfer, and the
extremes of stellar magnetic activity.  They are comprised of an
early-type main-sequence primary and a less massive late-type subgiant
secondary that fills its Roche lobe and loses mass to the
primary.\footnote{I adopt the common convention in which the primary
component refers to the currently more massive star of an Algol-type
binary, and the hotter (and usually less massive) star of an
RS~CVn-type binary.}  The secondary is the initially more massive star
of the system that has evolved first, but has then been demoted by
transfer of its mass onto the initially less massive primary.  The
comparatively faint mass-losing secondary can be stripped down to
layers whose composition has been radically altered by core hydrogen
burning (e.g.\ Sarna \& De~Greve~1996).

The late-type secondaries of Algol systems are strongly
magnetically active.  Tidal forces synchronise rotational and orbital
periods of these close binaries, and consequent rapid rotation excites
dynamo action in the convection zone of the evolved star that is
manifest at the surface in the form of chromospheric and coronal
emission (eg S.A.~Drake, Simon \& Linsky 1989; Singh, S.A.~Drake \&
White 1995).  Radio and X-ray activity levels of Algol systems are
similar to, but slightly lower than, those of their close relatives
the short-period RS~CVn-type binaries, in which both members are
late-type stars (Singh et al.\ 1996; Sarna, Yerli \& Muslimov 1998).


Eclipsing Algol-type binaries provide very accurate stellar masses and
radii with which to confront sophisticated models that follow the
evolution of the component stars, including Roche lobe overflow and
mass transfer (e.g.\ Sarna 1992,1993; Nelson \& Eggleton 2001).
Quantitative determination of element abundances modified by nuclear
processing in stellar interiors can provide critical tests of these
evolutionary models.  In particular, the degree of C depletion and N
enhancement through CN-cycle processing in the secondary can provide a
measure of the depth to which its outer layers have been stripped.  In
turn, the primary receives mass that might have been partially
processed by nuclear burning but that is then diluted by
rotationally-induced and thermohaline mixing with its pristine envelope.
Measurements of C and N in the primary can then help constrain these
mixing processes and the initial stellar mass of unprocessed material.

Based on high-resolution optical spectra, Parthasarathy, Lambert \&
Tomkin (1983) showed that the secondaries of the Algol binaries U~Cep
and U~Sge were depleted in C and enhanced in N by significantly
greater amounts than field giants of similar spectral type.  The more
pronounced abundance changes are the result of the CN-processed
material being mixed with a less massive outer envelope, confirming
that these stars lost a significant amount of mass prior to the onset
of deep convection.  Cugier \& Hardorp (1988), Cugier (1989), Dobias
\& Plavek (1985) and Polidan \& Wade (1992) also found C depletions in
the primaries and secondaries of several systems based on
International Ultraviolet Explorer (IUE) observations of C~II and C~IV
absorption and emission lines.  Peters \& Polidan (1984) found
evidence for C depletion in circumstellar matter in several systems,
and an extreme C depletion was found in AU~Mon.  Further studies of C
and N abundances based on optical spectra of several Algol systems
undertaken by Tomkin \& Lambert (1989), Yoon \& Honeycutt (1992),
Tomkin, Lambert \& Lemke (1993) and Tomkin \& Lambert (1994) confirmed
the general pattern of moderate to strong C and N abundance changes in
secondary stars and less extreme changes in the primaries.  The
general picture of the chemical evolution of Algols and observational
results supporting it has been summarised and explained in the review
of Sarna \& DeGreve (1996).

Algol primaries dominate the UV and optical spectra of these systems
and are generally accessible for abundance studies.  However, the
comparatively faint secondaries can only be studied reliably during
total eclipse.  Moreover, the rapid rotation of these stars smears out
spectral lines rendering high resolution spectroscopic studies more
uncertain.  In the case of Algol itself, no C and N abundances results
exist for the secondary star.

In contrast to UV-optical wavelengths, at X-ray wavelengths Algol
systems with primaries of spectral type mid-B to late-A are entirely
dominated by the magnetically active secondary.  Primaries in this
spectral type range do not have either an outer convective envelope
thought to be required to sustain strong coronal activity, or a
strong, shocked stellar wind thought to be responsible for X-ray
emission from O and early B stars.


In this paper, using {\it Chandra} Low Energy Transmission Grating
spectrograph (LETGS) X-ray spectra of Ly$\alpha$ resonance transitions
in C and N H-like ions (\S\ref{s:obs}), I determine for the first time
C and N abundances in the secondary of Algol (\S\ref{s:abun_rats}).
This study is performed relative to a ``standard'' star, the RS~CVn
system HR~1099.  As this work was nearing completion, Schmitt \& Ness
(2002) presented an analysis of C and N lines in several different
stars, including Algol.  Though they did not derive formal values for
C and N abundances, their work demonstrated that the surface 
abundances of the secondary of Algol have indeed been strongly
modified through CN-processing.  Ness et al.\ (2002) have also
recently presented an analysis of other aspects of the {\it Chandra}
LETG spectrum, including electron densities and temperature diagnostics.

In \S\ref{s:theo_comp}, I compare the
abundances derived here with theoretical predictions and show that
these are consistent with current evolutionary models for Algol.
Uncertainties and possible complications for this analysis are
discussed in \S\ref{s:uncert}.  The X-ray deficiencies of Algol
systems compared to RS~CVn-type binaries are discussed briefly in
\S\ref{s:comment}, and a summary of this paper is presented in
\S\ref{s:summary}.

\section{Observations and Analysis}
\label{s:obs}

Chandra LETGS spectra of Algol and HR~1099 where obtained in 2000
March and 2001 January, respectively, using the LETG and High
Resolution Camera spectroscopic (HRC-S) detector in its standard
instrument configuration.  The observations are summarised in
Table~\ref{t:obs}.  Observational data were obtained from the public
Chandra Data Archive\footnote{http://asc.harvard.edu/cda}.
Pipeline-processed (CXC software version 6.3.1) photon event lists
were reduced using the CIAO software package version 2.2, and were
analysed using the PINTofALE IDL\footnote{Interactive Data Language,
Research Systems Inc.} software suite (Kashyap \& Drake 2000).
Processing included filtering of events based on observed signal
pulse-heights to reduce background (Wargelin et al., in preparation).
The spectra of the two stars are compared throughout the wavelength
range of the LETG+HRC-S in Figure~\ref{f:spectra}, and in the limited
wavelength range 20-38~\AA\ in Figure~\ref{f:cno}; identifications for
prominent spectral lines are also indicated.

\begin{figure}
\plotone{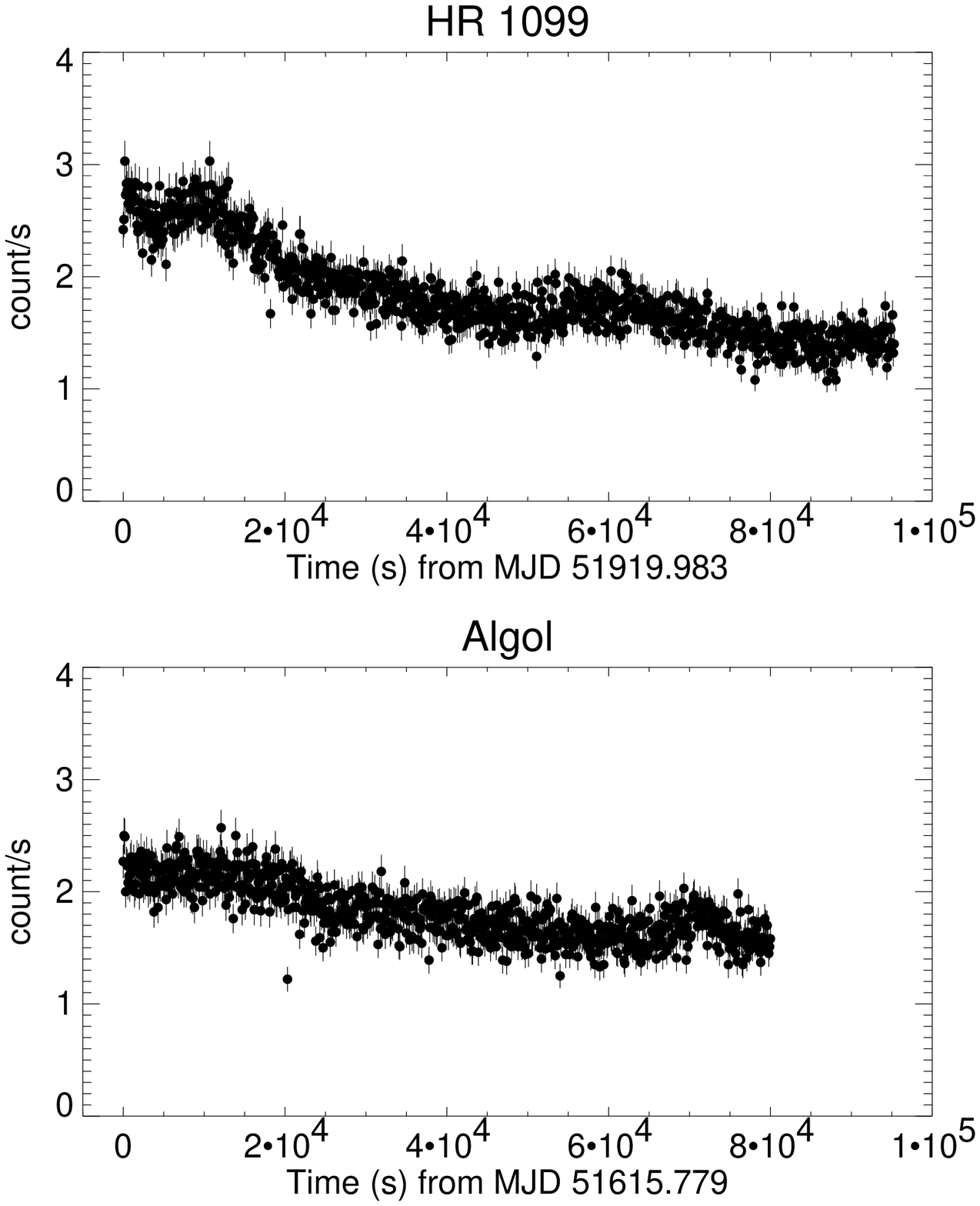}
\caption{{\it Chandra} LETG+HRC-S X-ray light curves of HR~1099 (top)
and Algol (bottom) obtained from 0th order events.  Each
light curve is binned at 100s intervals.  Based on the ephemeris of
Kim (1989), The Algol light curve spans
phases $\phi=0.74$ to 0.07.
}
\label{f:lc}
\end{figure}

\begin{figure}
\epsscale{0.4}
\plotone{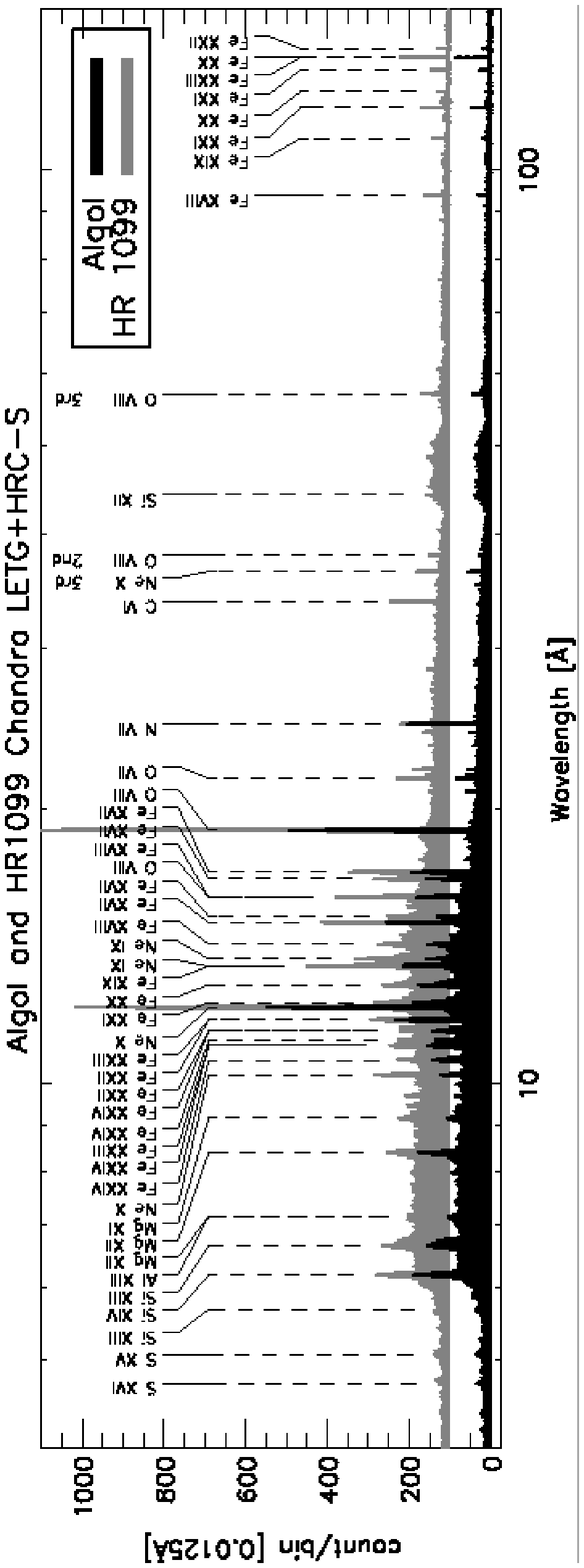}
\caption{
Illustration of the remarkable similarity in the combined positive and
negative order LETG+HRC-S spectra of 
Algol and HR1099 binned at 0.0125~\AA\ intervals.}
\label{f:spectra}
\end{figure}

\begin{figure}
\epsscale{1.0}
\plotone{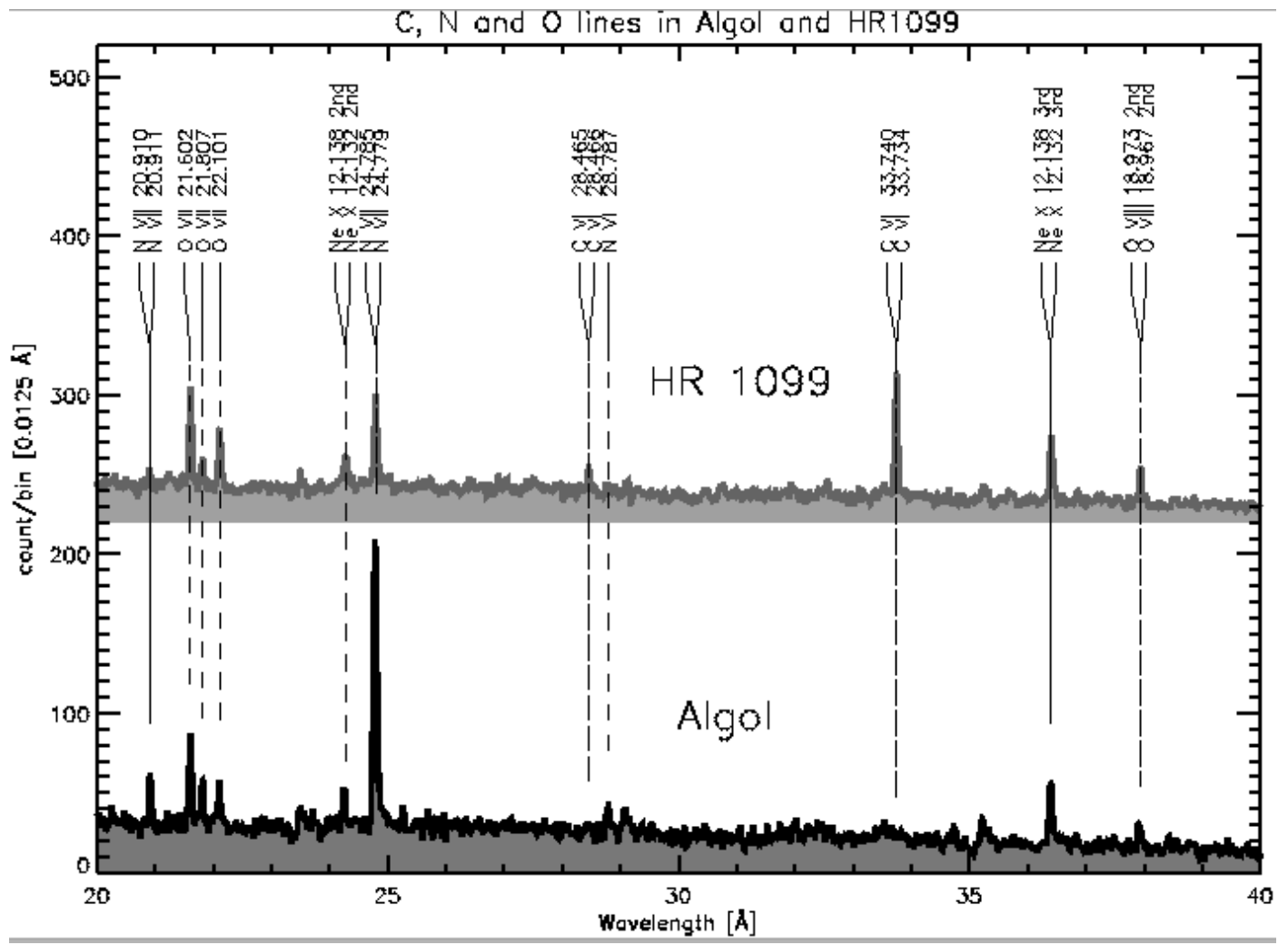}
\caption{
Comparison of the 20-35~\AA\ range containing spectral lines due to
C, N, O and Ne in Algol and HR~1099 LETG+HRC-S spectra.}
\label{f:cno}
\end{figure}

I first examined each data set in order to determine the extent to
which HR~1099 and Algol might have varied during the observations.
This is important because it is suspected that large flares are
accompanied in some cases by detectable changes in the coronal
abundances of the plasma that dominates disk-averaged spectra (see,
eg, Drake 2002 for a recent review).  Light curves for both
observations were derived using the 0th order events extracted in a
circular aperture that were then binned at 100s intervals.  These
light curves are illustrated in Figure~\ref{f:lc}.  Both are smooth
and devoid of obvious flare activity.  I therefore conclude that these
observations are representative of HR~1099 and Algol during times of
quiescence, and are thus well-suited to comparing the coronal
abundances of these stars.  It is also worth noting the orbital phase
during the observations: according to the ephemeris of Kim (1989), the
observation started at phase $\phi=0.74$ and ended at $\phi=0.07$, where
$\phi=0$ when Algol~B star is nearest the observer.  No significant
spectral evolution was seen during this period, consistent with an
earlier HETG observation analysed by Chung et al.\ (in preparation).
Irradiation by the strong UV flux of the
primary does not then have a significant influence on the gross
features of the spectrum, though radiative excitation of transitions
from the metastable levels in He-like ions has been observed (e.g.\
Ness et al.\ 2002).

I restrict the following spectral analysis to the wavelength range 15-35~\AA,
which contains the prominent lines of C, N and O, and also includes
the prominent 2p$^5$3d~$^1$P$_1$ $\rightarrow$ 2p$^6$~$^1$S$_0$ 
resonance transition of Fe~XVII at 15.01~\AA\ and H-like
Ne Ly$\alpha$ seen in second and third orders.  Spectral line fluxes
were measured by fitting ``modified Lorentzian'' functions of the form
$F(\lambda)=a/(1+\frac{\lambda-\lambda_0}{\Gamma})^\beta$, where $a$
is the amplitude and $\Gamma$ a characteristic line width.  For a
value of $\beta=2.4$, it has
been found that this function represents the
line response function of the LETG+HRC-S instrument to the photometric
accuracy of lines with of order a few 1000 counts or less (Drake et
al., in preparation).  The apparent continuum level in the spectral
region 25.5-28.0~\AA, judged by both visual inspection and examination
of radiative loss model spectral line lists to be ``line-free'', was
also determined in order to obtain a relative measure of continuum
flux intensity.

The flux, $F_{ji}$, from a collisionally-excited transition
$j\rightarrow i$ in an ion of an element with abundance $A$ in an
optically-thin collision-dominated plasma can be written as 
\begin{equation}
F_{ji} = AK_{ji} \int_{\Delta T_{ji}} G_{ji}(T) 
\Phi[N_e(T),V(T)] \;dT
\,\,\, \mbox{erg cm$^{-2}$ s$^{-1}$}
\label{e:flux}
\end{equation}
where $K_{ji}$ is a known constant which includes the frequency of the
transition and the stellar distance, and $G_{ji}(T)$ is the
``contribution'' function of the line that describes the temperature
dependence of the line emissivity.  This quantity defines the
temperature interval, $\Delta T_{ji}$, over which the line is formed,
and is dependent on the atomic physics of the particular transition.
The kernel $\Phi[(N_e(T),V(T)]$ is commonly known as the {\em
differential emission measure} (DEM)
\begin{equation}
\Phi[N_e(T),V(T)]=\overline{N_e^2}(T)\frac{dV(T)}{dT}
\label{e:dem}
\end{equation}
and describes the temperature structure of the excitation power of the
plasma, which is proportional to the mean of the square of the
electron density, $N_e$, and the emitting volume $V$.

If one makes the assumption that the DEMs of the coronae of Algol and
HR~1099 are very similar in terms of temperature dependence, then the
expression for the ratio of fluxes in a given transition observed from
the two stars reduces to the product of the ratio of abundances, $A$,
and the ratio of the emission measures of the two stars,
\begin{equation}
\frac{F_A}{F_H}=\frac{\Phi[N_e(T),V(T)]_A}{\Phi[N_e(T),V(T)]_H} 
\frac{A_A}{A_H}=Q\frac{A_A}{A_H}
\label{e:flux_rat}
\end{equation}
where the subscripts $A$ and $H$ denote Algol and HR~1099,
respectively, and $Q$ represents the ratio of source DEMs.
I adopt that assumption for the analysis here; I will further attempt
to justify this in \S\ref{s:abun_rats} and comment on its
propriety in \S\ref{s:uncert}.  

The ratio $Q$ can be determined from the 
source continua.  The continuum flux {\em density} $F_c(\lambda)$ per
unit wavelength interval can be approximated by the integral over
temperature of the product of the DEM kernel and the expression for
the emissivity of continuum processes (e.g.\ Mewe, Lemen \&
van den Oord 1986)
\begin{equation}
F_c(\lambda)= K_c\int_T \frac{G}{\lambda^{2} T^{1/2}}
\exp \frac{-143.9}{\lambda T}\Phi[N_e(T),V(T)] \;dT 
\;\; {\rm erg\; cm^{-2}\; s^{-1}\; \AA^{-1}}
\label{e:cont}
\end{equation}
where $K_c$ is again a constant involving the stellar distance, 
and $G$ is the total Gaunt factor
representing the sum of the Gaunt factors for the free-free,
free-bound and two-photon processes for all elements in the source
plasma.  In practice, the total Gaunt factor is dominated by that of
H, and to a lesser extent, He.  Provided the temperature dependence of
the DEM is approximately the same in both stars, the ratio of DEMs, 
$Q$, reduces simply to the integrated 
continuum flux ratios within an arbitrary wavelength interval:
\begin{equation}
Q=\frac{F_{cA}}{F_{cH}}
\label{e:cont_rat}
\end{equation}


Observed intensities (total counts) and their ratios, $F_{A}/F_{H}$,
for the features of interest here are listed in Table~\ref{t:fluxes}.
Note that no lines of C were detected in the Algol spectrum, and only
an upper limit could be obtained for H-like Ly$\alpha$~33.74~\AA.  
This is dramatically illustrated in
Figure~\ref{f:cno} in which C~VI is clearly detected in HR~1099 in
$2p\rightarrow 1s$ and $3p\rightarrow 1s$ transitions.

\section{Abundance Ratios}
\label{s:abun_rats}
\subsection{Formal Results}

I first draw particular attention to the remarkable similarity
of the spectra of Algol and HR1099 illustrated in
Figure~\ref{f:spectra}.  The spectra show strong resemblence to one
another in both the distribution of lines due to ions of
different elements and formed at different coronal temperatures, and
in the relative strengths of these lines, throughout the LETG
range. It is clear from this comparison that the X-ray emitting
coronae of Algol and HR1099 essentially share the same gross
characteristics of temperature structure and relative element
abundances. As I show below, Carbon and Nitrogren abundances are
conspicuous exceptions.

A further test that the temperature structures of the DEMs of Algol
and HR~1099 are very similar, at least in the temperature range that
concerns us for the elements C and N,  can be found in the
temperature-sensitive ratios of H-like to He-like transitions among
the CNO trio.  While the He-like complexes of both N and C are both very
weak and
blended in Algol and HR~1099 spectra, the He-like O~VII resonance line
$1s2p\, ^1P_1 \rightarrow 1s^2\, ^1S_0$ is well-observed and can be
compared with the H-like $2p\,  ^2P_{3/2,1/2} \rightarrow 1s\, 
^2S_{1/2}$.  The observed photon counts in each line are listed for
the two stars in Table~\ref{t:fluxes} and their ratios 
$F_{OVIII}/F_{OVII}$ are $9.9\pm 1.2$ and $10.1\pm 1.0$
for Algol and HR~1099, respectively---identical to within experimental
uncertainties.

Turning now to C and N, a qualitative sense of the 
abundances of these elements in Algol relative to HR~1099 is
immediately apparent on inspection of Figure~\ref{f:cno}.  Lines of
H-like N are stronger by more than a factor of two in Algol, whereas C
is strong in HR~1099 yet undetected in Algol.  Similar results were
presented by Schmitt \& Ness (2002).

Relative element abundances were derived from the observed
line and continuum fluxes using 
Equations~\ref{e:flux}-\ref{e:cont} in \S\ref{s:obs}.  
I express abundances in Algol
relative to those in HR~1099, thereby using the latter as the
``standard'' star.  In the conventional spectroscopic bracket
notation, the logarithmic relative abundance of element X,
[X/H]$_{1099}$, is given by the logarithmic difference between line
and continuum flux ratios:
\begin{equation}
\mbox{[X/H]}_{1099}=\log_{10}\frac{F_{XA}}{F_{XH}} - 
\log_{10}\frac{F_{cA}}{F_{cH}}
\label{e:rel_abun}
\end{equation}
The resulting abundances for Fe, O and Ne are listed in
Table~\ref{t:abuns}.  The corresponding numbers for C and N are 
[C/H]$_{1099}<-1.13\, (3\sigma)$ and  [N/H]$_{1099}=0.31$.  The formal
uncertainty on the N abundance based on the line intensity
measurements is only 0.02~dex, though a more
realistic uncertainty is of order 0.1~dex (\S\ref{s:uncert}).
These C and N abundances demonstrate formally the severe depletion of C and
enhancement of N in the corona of Algol B.  Expressed as an abundance
ratio, I obtain [C/N]$_{1099}<-1.44$, which is a factor of 100 lower
than the solar photospheric value of C/N$=0.47$ (Allende Prieto et
al.\ 2002; Grevesse \& Sauval 1998).  

I do not quote the values [C/H]$_{1099}$ and [N/H]$_{1099}$ in
Table~\ref{t:abuns} because these quantities do not tell us directly
what the changes in C and N abundances resulting from nuclear
processing are.  To determine these changes, we need to ascertain the
values of the original C and N abundances in the unprocessed envelope
of Algol B.  There are two complications to this.  Firstly, as noted
in \S\ref{s:intro}, coronal abundances in at least some stars appear not to
reflect the values in the underlying stellar photosphere.  Secondly, and
again as noted in \S\ref{s:intro}, both Algol and HR~1099 are
``metal-deficient'' in their coronae with respect to the {\em solar
photosphere}.   I address this latter point first.

\subsection{Establishing the Photospheric Baseline}

Transitions involving electrons with ground states in the Fe $n=2$
shell tend to dominate soft X-ray spectra of hot coronal plasmas, and
for this reason Fe is often used as a proxy for the global plasma
metallicity.  The coronal Fe abundances for Algol and HR~1099 are identical to
within observational uncertainty.  A recent measurement of the
absolute Fe abundance in the corona of HR~1099, Fe/H$=7.0\pm 0.1$
(corresponding to [Fe/H]$=-0.5$ relative to the solar photospheric
abundance; e.g.\ Grevesse \& Sauval 1998) was recently presented by
Drake et al.\ (2001).  Earlier estimates of the coronal Fe abundance
of Algol have been made based on ASCA ([Fe/H]$=-0.32\pm 0.01$, with
the error based on statistical uncertainty only; Antunes, Nagase \&
White 1994; see also Singh, Drake \& White 1996) EUVE
([Fe/H]$=-0.23$ to $-0.66$; Stern et al.\ 1995), while Favata \& Schmitt
(1999) obtained a global metallicity of [M/H]$\simeq -0.35$ during
quiescence in an observation by BeppoSAX that captured a very large
flare (the metallicity was interpreted as undergoing strong
variations correlated with the flare rise and decay).  

The determination of the coronal Fe abundance of Algol, ``bootstrapped''
from the earlier Drake et al.\ (2001) absolute Fe abundance of
HR~1099, is in fairly good agreement with these values, and I
consider this result quite robust.
In quoting these other abundance results, 
I note that the works cited all adopted the
solar abundance compilation of Anders \& Grevesse (1989), which
employed an iron abundance Fe/H$=7.67$.  I have accordingly adjusted the
relative abundances to a solar iron abundance of Fe/H$=7.50$ (see
Drake, Laming \& Widing 1995 for further details) used by Drake et
al.\ (2001).

Having established the Fe abundance in both Algol and HR~1099, the
next question is whether or not the value obtained is representative
of their photospheric Fe abundances, or whether the coronal abundances
have been altered by fractionation processes.  To my knowledge, no
detailed photospheric abundance analyses of either Algol A or B have been
published and so a definitive answer to this question will have to
wait.  In either case, we have to consider what the initial C and N
abundances, prior to mixing of the envelope and nuclear-processed
core, would have been.

The elements C and N do not share the same nucleosynthetic origins:
the dominant isotope C$_{12}$ is produced mostly by He-burning, while the
dominant N isotope N$_{14}$ is produced mainly by H-burning. 
Studies of C and N photospheric abundances in stars of
different metallicity (as represented by Fe) show different trends.
The abundance of C exhibits moderate enrichment relative to Fe with
decreasing [Fe/H], amounting to [C/Fe]$\sim +0.2$ at [Fe/H]$\sim -0.7$
(e.g. Tomkin, Woolf \& Lambert 1995, Gustafsson et al.\ 1999; Shi,
Zhao \& Chen 2002).  In contrast, the abundance of N appears to follow
that of Fe over a range of [Fe/H] from $-0.8$ to +0.1 (e.g.\ Shi et
al.\ 2002).  If Algol and HR~1099 were mildly metal-deficient, and
this reflected the actual stellar compositions and was 
not the result of coronal compositional fractionation, then
we might therefore expect original C and N abundances of [C/H]$\simeq
-0.35$ and [N/H]$\simeq -0.5$.  We can use these values, together with
the C, and N solar photospheric abundances C/H=8.39 (Allended Prieto
et al.\ 2002),  N/H=7.92 (Grevesse \& Sauval 1998)  
to estimate projected changes in the
number of C and N atoms in the envelope of Algol~B.  In the
usual logarithmic notation for the number density relative to
hydrogen, these are $\Delta$C/H$\la -8.0$ and $\Delta$N/H$\simeq
+7.4$.  These values pose an immediate problem: in the CN-cycle, N is
produced at the expense of a commensurate depletion in C, and the total
number of C and N nuclei is expected to be conserved.  That we do not
arrive at equal changes in C and N implies that the initial
unprocessed abundances assumed here are incorrect.  This conclusion is not 
changed significantly because of the adoption of a slightly non-solar 
initial C/N abundance ratio expected in a disk star with [Fe/H]$=-0.5$; 
the difference only amounts to C/N being larger by 0.15~dex.

\begin{figure}
\plotone{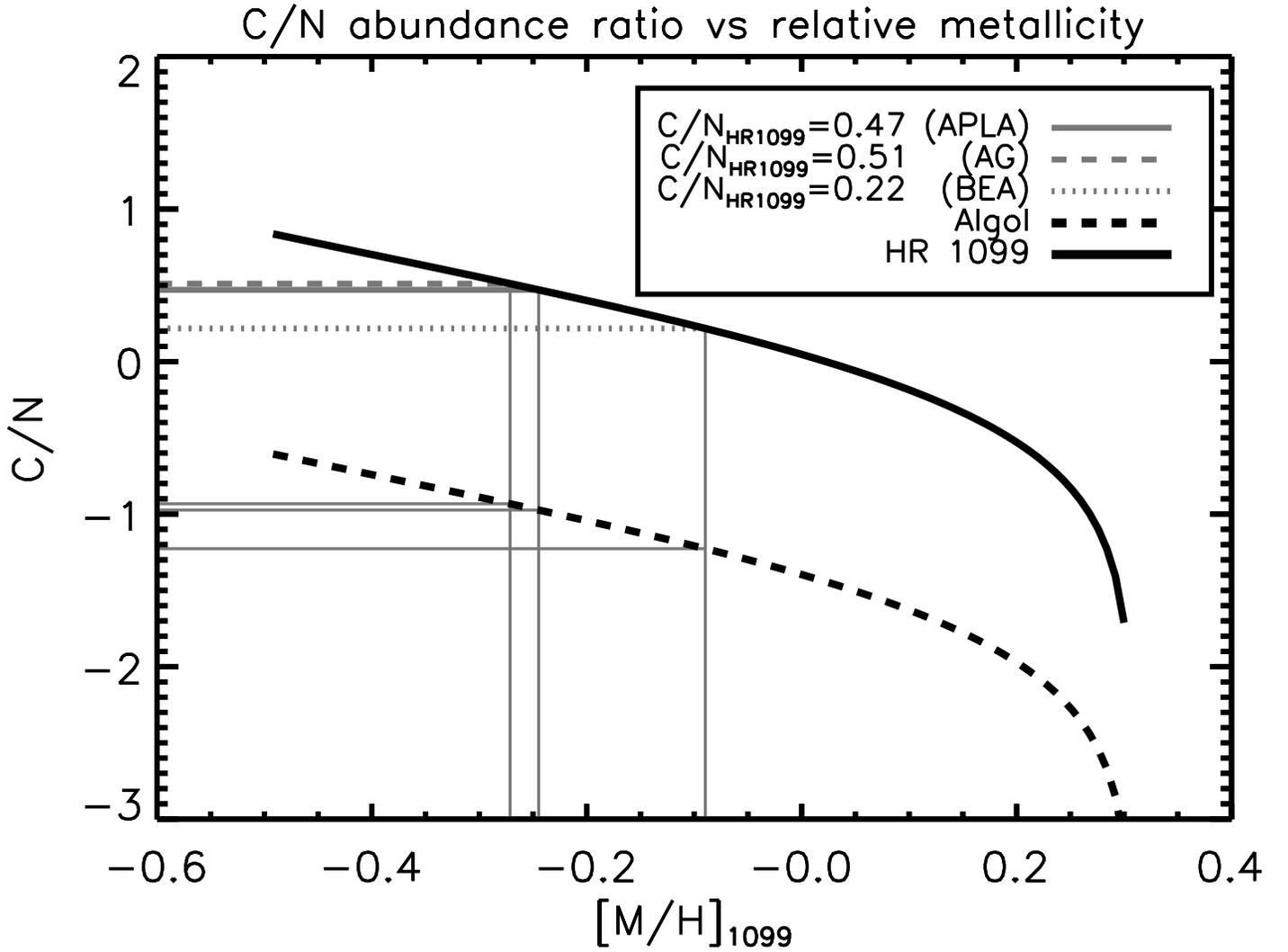}
\caption{
The derived C/N abundance ratios by number in Algol and HR~1099 as a
function of the logarithmic metallicity difference between the two
stars ([M/H]$_{1099}=\log_{10}(n(M)_A/n(M)_H)$; see also
Equation~\ref{e:totcn}).  Unprocessed C/N ratios in HR~1099
corresponding to solar photospheric 
C and N abundances from Allende Prieto et al.\
(2001,2002; C abundance combined with the N abundance from Grevesse
\& Sauval 1998; APAL), Anders \& Grevesse (1989; AG), and to the estimates
for the coronae of HR~1099 derived by Brinkman et al.\ (2001; BEA), are
indicated by the thicker horizontal straight lines.  The C/N ratio for
Algol and the metallicity parameter corresponding to these C/N ratios
in HR~1099 are indicated by the thinner solid lines.
}
\label{f:cn}
\end{figure}

An alternative approach is to determine what initial absolute 
abundances of C and N would match the observed C depletion and N enhancement 
in Algol relative to HR~1099.  Or, put another way, what value of
metallicity, [M/H], would be required for the total number of C and N
nuclei to be conserved.  The number density of an element $n(X)$ is
given by our observed fluxes and by Equations~\ref{e:flux_rat} and
\ref{e:cont_rat}.  We write the observed 
ratios of C and N in Algol and HR~1099 as
\begin{equation}
\frac{n(C)_{A}}{n(C)_{H}}=\delta C \;\;\;\;\; \mbox{and} \;\;\; \;\;
\frac{n(N)_{A}}{n(N)_{H}}=\delta N
\end{equation}
We now assume, firstly, that the total number density of C and N nuclei
in the processed envelope of Algol remains unchanged compared to the
initial pristine value, and, secondly, that this total number density
is the same in HR~1099, except for a scaling factor allowing for
different global metallicities between the two stars, 
\begin{equation}
n(C)_{A}+n(N)_{A}=m[n(C)_{H}+n(N)_{H}],
\label{e:totcn}
\end{equation}
where $m=n(M)_A/n(M)_H$ is ratio of metallicities by number.
The C/N abundance ratios by number for Algol and HR~1099 can then be written
\begin{equation}
\left(\frac{n(C)}{n(N)}\right)_A=\frac{m\delta C-\delta N}{\delta C\delta N
- m} \;\;\;\;\; \mbox{and} \;\;\;\;\; 
\left(\frac{n(C)}{n(N)}\right)_H=\frac{m-\delta N}{\delta C- m}
\end{equation}
I illustrate these C/N abundance ratios as a function of the
logarithm of the metallicity parameter $m$ in Figure~\ref{f:cn},
adopting the derived C abundance upper limit for Algol as a proxy for
use in the value $\delta C$.   
Also indicated on this figure are the C/N abundance ratios from three
sources: the solar ratio derived from the C abundance of Allende
Prieto et al.\ (2002) and the N abundance of Grevesse \& Sauval (1998)
that I adopt as ``reference'' solar values; the ratio from the still
commonly used solar abundance compilation of Anders \& Grevesse
(1989); and  the ratio from the abundance values derived relative to
that of O for the coronae of HR~1099 itself based on 
XMM-Newton observations by Brinkman et al.\ (2001).  The intersection
of these C/N ratios with the HR~1099 C/N locus indicates the 
metallicity of Algol relative to HR~1099 needed in order that the
number density of C and N abundances be conserved.  For the solar
C/N ratios, we need a metallicity [M/H]$_{1099}\sim -0.25$, whereas
for the Brinkman et al.\ (2001) C/N ratio, the metallicity is
$-0.1$~dex.  The uncertainties of the values are of order 0.1~dex or
so.  Note that the metallicity derived in this way is not sensitive to
the fact that I have used the {\em upper limit} to the C ratio,
$\delta C$: nearly all (over 90\%) of the C in Algol has already 
been converted into N, and so the N abundance remains essentially the same
regardless of the exact C abundance adopted.

I have already established above that the Fe abundances in the coronae
of Algol and HR~1099 are essentially the same.  The ``metallicity'' values
indicated by the C/N ratios suggest slightly lower abundances in Algol than
HR~1099, and possibly significantly so.  This ``metallicity'' is of course
just a number representing C and N.  However, I note that the 
abundances for O and Ne are similar to this in being slightly lower in
Algol than in HR~1099; the error-weighted mean of the Ne and O abundances
(Table~\ref{t:abuns}) yields a ``metallicity'' [M/H]$_{1099}=-0.20$, a
value consistent, within errors of 0.1~dex or so, with the C/N result.
I take this agreement as indicating that indeed the light element
abundances in Algol are lower by of order 0.2~dex than those in
HR~1099.  

By assuming an initial C/N ratio and that this ratio is represented by
the current C/N ratio in HR~1099, we arrive at a C/N ratio for Algol;
these ratios are indicated in Figure~\ref{f:cn} and are C/N=$-0.95$
for initial solar C/N ratios, and C/N=$-1.2$ for an initial ratio
equal to that of Brinkman et al.\ (2001).


\section{Comparison with Theoretical Models}
\label{s:theo_comp}

As discussed above in \S\ref{s:abun_rats}, the assumption that the
present-day C/N coronal abundance ratio in HR~1099 is representative
of the ratio in the original unprocessed envelope of Algol~B, and that
this ratio in HR~1099 is similar to the solar ratio, implies
the change in the C/N abundance ratio in Algol~B as a result of
nuclear processing and dredge-up is $\Delta\log C/N \leq -1.44$~dex.
In comparison, the changes expected at the surface of normal G and K
giants in the post-dredge-up phase is $\Delta\log C/N \sim -0.3$ to
$-0.7$ (eg Lambert \& Ries 1977, 1981; see also \S\ref{s:two-zone} below).
Main-sequence CN-processing in the core of Algol~B is expected to have
been essentially the same as that in a single star of the same
mass; the extreme decrease in the post-dredge-up phase in Algol~B is
simply a result of a lower dilution factor of this CN-processed
material because of the loss of material from its outer envelope to Algol~A.

\subsection{Close Binary Evolutionary Predictions}

Stellar evolution calculations for Algol-type binaries  that follow the
surface abundances of C and N have been undertaken for intermediate mass
systems by de~Greve
(1989,1993; see also de~Greve 1986), and Sarna (1992,1993).  These
authors conclude that the mass transfer in Algol corresponds to early
case~B, in which transfer is initiated by the expanding stellar
envelope filling its Roche Lobe after turn-off from the main-sequence and the 
establishment of hydrogen shell burning.  In the early stages of mass
transfer, the convection zone of the initially more massive Algol~B 
would not have been fully-developed and so the envelope from which
material was lost would only be
partially mixed with deeper layers that had been subjected to
CN-processing.  During later stages, after most of the pristine outer
envelope had been lost to Algol~A, material from layers in which
CN-processing had reached equilibrium, and that would also have been
He-rich from H-burning, was transferred.  Being He-rich, this material
would have induced thermohaline mixing on Algol~A (mixing resulting
from the larger molecular weight of the overlying material), diluting
the CN-processed material with the rest of the envelope of Algol~A.

I compare the abundance results derived here with the detailed binary
model evolutionary calculations specifically for Algol performed by
Sarna (1993).  These models include the effects of non-conservative
mass transfer, and mass and angular momentum loss through magnetic
activity and stellar winds---effects that have been shown to be
important for intermediate and lower mass Algols that include a
magnetically-active late-type component by, e.g., Sarna et al.\
(1997), Eggleton (2000) and Nelson \& Eggleton (2001).  
Sarna (1993) was able to derive a
model that successfully matched the present day observed parameters of
the system, including the C abundance in Algol A obtained by Cugier \&
Hardorp (1988) based on C~II lines observed by the International
Ultraviolet Explorer (IUE).  The C abundance appears depleted in Algol
A by about a factor of 2 because of the dilution of its original
envelope by material strongly depleted in C from Algol B.  Changes in
C and N abundances in Algol B are predicted as part of these
calculations and amount to $\Delta\log C=-1.0$ (Sarna 1993) and
$\Delta\log N=+0.4$ (M.~Sarna, private communication), yielding a
change in the C/N abundance ratio of $\Delta\log C/N =-1.4$.

Within reasonable uncertainties, the Sarna (1993) abundances
are quite consistent with the values derived in this study.  As
discussed at length above, the individual C and N abundances depend on
the adoption of the metallicity for these elements.  Formally, then,
our adopted N abundance is 0.1~dex higher than the model prediction
and to lower it by this amount requires a lower C abundance of
$-1.03$.  Since our C abundance is a $3\sigma$ upper limit, it does
seem possible that the C/N decrease could be significantly greater
than $-1.4$~dex predicted by the Sarna (1993) best-fitting
model and it might be of interest to determine what changes to this
model might accommodate a larger depletion.  However, at the present
day phase in its evolution, this model also predicts quite a steep
decline in the C abundance and it seems plausible that a  
greater depletion in the
C/N ratio might be accommodated by the model uncertainties.

\subsection{Simple Two-Zone Model}
\label{s:two-zone}

It is worth comparing the C/N ratio we have obtained for Algol with
the values expected in single G and K giants following convective
dredge-up of CN-processed material during the ascent of the red giant
branch.  Lambert \& Ries (1977,1981) obtained typical values by number
of $n(C)/n(N)\sim 1$-2 for a sample of field giants, to be compared
with their adopted solar (initial) ratio of 4.8, or a logarithmic
change in the C/N ratio of $\Delta\log C/N \sim -0.3$ to $-0.7$.  These
values were 
in good agreement with the evolutionary models of Dearborn, Tinsley \&
Schramm (1978), which, for a star of mass $2.8~M_\odot$ and the same
initial C/N ratio, predicted a surface abundance ratio of 
[C/N]$=-0.55$ after dredge-up.  

Following Parthasarathy, Lambert \& Tomkin (1983), by adopting a crude
two-zone model for the envelope and CN-processed shell we can make a
very rough estimate of the fraction of mass lost from the star based
on the (assumed) initial abundances and (measured) final abundances.
In this model, one assumes that mass transferred onto Algol~A is
comprised of only pristine unprocessed material prior to deep mixing
of envelope and processed core.  At dredge-up, a remaining outer
envelope of mass $M_u$ containing only unprocessed material is mixed
with a mass $M_p$ of matter processed through the CN-cycle on the
main-sequence.  We can then equate the number of atoms of C or N per
unit mass in the mixed material, $n_m$, to the numbers per unit mass
in the processed core, $n_p$, and unprocessed envelope, $n_u$, as
follows:
\begin{equation}
n_m(M_u+M_p)=n_u M_u+n_p M_p
\end{equation}
The ratio of the processed and unprocess masses is then simply
\begin{equation}
\frac{M_p}{M_u}=\frac{n_u-n_m}{n_m-n_p}
\label{e:mass_rat}
\end{equation}
Since we have only an upper limit for the C abundance $n_m$, I
estimate $M_p/M_u$ based on the observed N abundance.  As before, I
adopt the initial solar C abundance of Allende-Prieto et al.\ (2002)
and the N abundance of Grevesse \& Sauval (1998).  The observed N
abundance is then [N/H]$=-0.56$, where the value in
Table~\ref{t:abuns} has been lowered by 0.05~dex to be in accordance
with the metallicity [M/H]$=-0.25$ indicated for the adopted solar C
and N abundances in Figure~\ref{f:cn}. Assuming that in the processed
mass $M_p$ essentially all the C is converted to N, the mass ratio
given by Equation~\ref{e:mass_rat} is $M_p/M_u=8.2$.  

As pointed out by Parthasarathy et al.\ (1983) based on the work of
Iben (1967), in the absence of strong mass loss, and over a fairly
wide range of stellar masses, standard evolutionary models for
intermediate mass stars such as the Algol~B progenitor have an
unprocessed envelope mass fraction of about half the total stellar
mass, or $M_p/M_u\sim 1$.  The simple two-zone model then provides a
``0th order'' proof that Algol~B must have lost nearly all of its
outer envelope, or about half its mass.  How applicable is such a
simple two-zone model to Algol?  It implicitly 
assumes that mass loss occurred only during
a rapid phase that was short compared to evolutionary timescales.
Empirically, this cannot be the case because the C abundance of
Algol~A indicates that He-rich material that induced thermohaline
mixing must have been transferred,
otherwise the surface abundances of the two stars would be about the same.
Such He-rich layers in Algol~B also correspond to layers in which C
and N abundances would have reached the equilibrium values of
CN-processing, where C is reduced to about 3\% of its initial value
([C/H]$\sim -1.5$).  The C abundance is larger than this equilibrium
value because of mixing of the pristine envelope with the CN-processed
layer during mass-transfer.  The two-zone model estimate is therefore
a lower limit to the actual mass loss.  The model of Sarna (1993)
suggests an initial mass of Algol~B of $2.81 M_\odot$, to be compared
with its present mass of $0.81 M_\odot$; in this non-conservative
model, 15~\%\ of the total mass is lost to the system.  

Prior to the Chandra LETGS observations, the only direct
nucleosynthetic observational clue as to the past evolution of the
Algol system was the C abundance estimate for Algol A of Cugier \&
Hardorp (1988).  As pointed out by Sarna (1993), this value remained
too uncertain to be of quantitative value in constraining evolutionary
models.  The Chandra observations present the first direct and
definitive observational evidence of the strong modification of the
surface C and N abundances of Algol B through exposure to
main-sequence nuclear burning and subsequent shedding of its outer
envelope.  These observations provide the best observational
confirmation of theoretical models for the evolutionary history and
present status of Algol.

\section{Further Consideration of Uncertainties}
\label{s:uncert}

The formal abundance results in Table~\ref{t:abuns} 
quote only statistical uncertainties, which are
very small owing to the use of prominent and well-observed spectral
features.  The advantage of the relative study adopted here is that
the uncertainties inherent in absolute abundance studies tend to cancel
each other out because they are the same for both stars.  These
include uncertainties in the instrument calibration, in the
measurement and extraction of line fluxes, and in the atomic data
relevant for the radiative emission of interest.

The largest uncertainties in this study instead arise from the two
main assumptions adopted (supplementary to the common underlying assumption
that the sources can be adequately described by collision-dominated,
optically thin plasmas, an assumption that we expect to be reasonably
accurate). These are: 
\begin{enumerate}
\item the temperature dependences of the source
DEMs are essentially the same; and 
\item there is no strong
differential compositional fractionation at work between the two
stars, especially among C and N, so that inferred abundance
differences represent underlying stellar compositional differences.
\end{enumerate}
These two assumptions are partly correlated: one might expect that any
compositional fractionation at work in one corona might be duplicated
in another that is in some respects very similar, provided the
fractionation mechanism(s) are at least partly dependent on the common
parameters.  I argue that this is indeed the case below.  Regarding
the temperature structures of the source plasmas, it has been found
that the ratios of He-like to H-like resonance lines in the abundant
elements O, Ne, Mg, Si, and S are very similar to within statistical
uncertainties (Drake et al.\ in preparation), with deviations
typically of order 20~\%\ or less.  These ratios are not
sensitive to abundances but are quite sensitive to temperature. 
Based on this evidence, provided assumption (2) is valid, we might
expect uncertainties in abundance ratios of order 0.1~dex to result
from small differences in source temperature structure.

\subsection{Similarities in Stellar Parameters and Coronae}
\label{s:similarities}

The principal stellar parameters currently thought to drive coronal
formation can be coarsely defined as spectral type and rotation rate.
In this regard, the similarity in spectral type of the active subgiant
primary in the HR~1099 system that dominates its coronal activity and
Algol B (Algol A being X-ray dark), together with their similar
rotation periods, suggests at the outset that their coronae might also
share very similar characteristics.  Further parameters for the X-ray
dominant components are summarised in Table~\ref{t:obs}.  These values
indicate that, indeed, the effective temperatures appear to be
consistent within experimental uncertainty, as, likely, are the masses
and radii.  These are known to an accuracy $\sim 5$~\%\ for Algol
(e.g.\ Richards et al.\ 1988), but for HR~1099 formal uncertainties
are difficult to estimate.  I have listed the Donati (1999) values of
these parameters which correspond to a system inclination $i=38\degr$;
this author notes the sensitivity of the derived values to the adopted
inclination, and that for an inclination similar to that of Fekel
(1983), both derived mass and radius are larger.  Also of note is the
temperature for Algol B derived by Kim (1989), of ``$4888\pm 96$~K'',
which is very similar to the HR~1099 effective temperature of Randich
et al.\ (1994).  Indeed, based on parameters derived, Kim (1989)
adopted a spectral type for Algol~B of K0~IV---very similar to the
K1~IV type of the primary of HR~1099.


I have not considered stellar metallicity here---a parameter that
could also influence coronal activity through alteration of convection
zone properties (e.g.\ Pizzolato et al.\ 2001), in addition to outer atmosphere radiative loss rates.
I address the stellar metallicities below; again these appear to be
similar in both stars.  A further parameter is the He abundance, which
is expected to be higher in Algol B owing to a smaller dilution of
H-burning products with an envelope depleted through mass transfer.  I
return to this below in \S\ref{s:comment}.

In summary, the remarkable similarity in the stellar parameters of
active Algol and HR~1099 components, as noted above, in addition to
their identical rotation rates, leads me to suggest that their coronae
are likely to be superficially identical.  This conjecture is borne
out in a detailed comparison of their X-ray spectra, in which
variations in relative line intensities are generally well within a
factor of two---see Figure~\ref{f:spectra}.  Similar conclusions were
also reached in a comparison of radio observations of Algol and
HR~1099 by Mutel et al.\ (1998).  They found that the mean radio
luminosity, luminosity distribution functions, and spectral index
correlations with flux density from Very Long Baseline Array and Green
Bank Interferometer (GBI) observations were nearly indistinguishable.
Some difference in circular polarisation fraction was attributed
naturally to differences in system inclination.  Some palpable
differences have been found, however: Richards, Waltman \& Ghigo
(2002) have recently presented evidence that apparent periodicities in
radio flaring in the time from 1995 to 2000 at frequencies of 2.3~GHz
and 8.3~GHz are different, with the strongest periodicities found
being $49\pm 5$ days and $120\pm 5$ days for Algol and HR~1099,
respectively.  Nevertheless, all other characteristics appearing very
similar, I 
feel that assumption (1) is justified and that, indeed, the
temperature structures of the coronae of HR~1099 and Algol are
essentially the same.

The similarity in the X-ray spectra of these stars also argues against
any significant role in coronal formation for their different Roche
lobe filling factors, or for the ongoing mass transfer in Algol, again
in agreement with the conclusions of Mutel et al.\ (1998) based on
radio properties.

\subsection{Coronal Abundance Anomalies and Compositional Fractionation}
\label{s:fractionate}

The possibility that the coronae of either, or both, Algol and HR~1099
have compositions that differ from their underlying photospheres would
affect this analysis and conclusions if there were to be a
relative fractionation between C and N that differed between the two
stars; ie, if the observed difference in the ratio C/N in the
coronae of Algol and HR~1099 were to be attributable to affects other
than purely CN-processing.  

As noted in \S\ref{s:intro}, both EUVE (Stern et al.\ 1995) and ASCA
(Antunes et al.\ 1994) studies of Algol obtained low values for its
coronal Fe abundance.  Both studies estimated [Fe/H]$\sim -0.5$, a
result that was confirmed by BeppoSAX observations reported by Favata
\& Schmitt (1999), and that is further supported here by our analysis
relative to HR~1099.  While a mild Fe deficiency of [Fe/H]$=-0.5$ is not
unusual for stars of the Galactic disk population, young systems
involving early-type stars are not generally expected to be so
metal-poor.  The best-fit evolutionary model of Sarna (1993) suggests
an age for Algol of about 0.45~Gyr (see also Nelson \& Eggleton 2001
whose best-fit model is 1.1~Gyr old).  The recent statistical
age-metallicity relationships for the Galactic disk of Feltzing,
Holmberg \& Hurley (2001) indicate that it would indeed be quite
unusual for stars earlier than F0 ($\ga 1.5M_\odot$) to have such low
Fe abundances.  Sarna (1993) and Nelson \& Eggleton (2001) Algol
models suggest Algol B was either a late B or early A zero-age
main-sequence star, with predicted masses of 2.8 and $2.2~M_\odot$,
respectively, confirming the statistical likelihood that Algol should
not be significantly metal-poor.  We conclude, then, that Fe is very
likely depleted in the corona of Algol B relative to the underlying
star.  

A recent review of coronal abundance studies for RS~CVn systems for
which estimates of both coronal and photospheric abundances exist also
supports the idea that Fe is in fact depleted in their coronae
relative to their underlying photospheres, possibly by as much as
1~dex (Drake 2002).  There are no definitive photospheric abundance 
studies that tell us whether or not this is also the case for HR~1099.
In addition, the evolutionary status of HR~1099 is more difficult to
infer than for Algol because derived stellar masses are dependent on
the adopted system inclination (e.g.\ Fekel 1983; Donati 1999).  If
the estimate of $\sim 1M_\odot$ for the primary listed in
Table~\ref{t:obs}, corresponding to an inclination $i\sim 40\degr$, is
correct, then its evolved state would imply an age of $\sim 10$~Gyr.
Such an old disk star would be expected on a statistical basis to be
somewhat metal-poor.

While no detailed photospheric abundance studies have been undertaken
for HR~1099, estimates of Fe abundances have been made as part of
other studies.  Randich et al.\ (1994) obtained [Fe/H]$=-0.6$ for the
primary in a study of the Li region near 6708~\AA, but obtained
[Fe/H]$=0$ for the secondary component, a metallicity pattern
apparently found also by Vogt \& Penrod (1982 private communication to
Fekel 1983).  Interestingly, Randich et al.\ (1994) derived similar
metallicity differences between primary and secondary components for
several other RS~CVn binaries in their sample.  Such a differences in
metallicity between components of a close binary are difficult to
understand on evolutionary grounds, but similar differences had been
suggested earlier based on spectroscopic analyses of a handful of both
Algol-type and RS~CVn-type binaries by Naftilan (1975) and Naftilan \&
Drake (1977,1980).  However, Ottmann et al.\ (1998) have refuted claims of
significant metal paucity for some systems in the Randich et al.\
(1994) sample based on more extensive spectroscopic analyses, while a
detailed study of the AR~Lac secondary by Gehren (1999) that obtained
[Fe/H]=0, in contrast to the result [Fe/H]$\sim -1$ of Naftilan \&
Drake (1977), casts doubt on the reality of true intra-binary photospheric
abundance differences.  More recently, in a Doppler imaging study of
HR~1099, Strassmeier \& Bartus (2000) estimated [Fe/H]$=-0.1$ and
[Ca/H]$=-0.2$ for the secondary of HR~1099.
 
It is difficult to draw firm conclusions regarding the true
photospheric metallicity of HR~1099 based on this summary of disparate
results.  Clearly, the issue of possible abundance differences
between primary and secondary components of close binaries warrants
further study: if true, such differences would provide a strong
challenge to structural and evolutionary models for such systems.  At
present, I appeal to the differential nature of this study and
tentatively suggest that HR~1099 is not significantly more metal-poor
than Algol, and that the very similar X-ray dominant components of
these systems are undergoing very similar fractionation processes that
results in Fe being depleted in their coronae by of order 0.5~dex.

In addition to this apparent depletion of Fe, the recent Chandra and
XMM-Newton studies of HR~1099 based on high resolution X-ray spectra
(Drake et al.\ 2001, Brinkman et al.\ 2001) indicate that other
element abundances are strongly modified in the corona.  In
particular, both studies found enhancements of Ne relative to Fe of
about a factor of 10, and of O relative to Fe of factors of 3-4.
Brinkman et al.\ (2001) argued that the differences followed an
inverse First Ionization Potential (FIP) effect---a situation opposite
to that in the solar corona where elements with FIP $\la 10$~eV appear
to be enhanced relative to elements with FIP $\ga 10$~eV.  Considering
the magnitudes of these abundance anomalies, the Ne and O abundances
are strikingly similar in the two systems, differing in their ratios
relative to Fe by only 0.2~dex.  Regardless of whether the
fractionation in these stars is FIP-related or otherwise, I conclude
that {\em the element fractionation processes at work in HR~1099 and
Algol~B are remarkably similar and produce essentially the same
abundance pattern}.  On this basis, I suggest that there is no
significant difference between the two stars in any coronal
fractionation at work between C and N.

Further support for this argument comes from the adopted C depletion
and N enhancement.  As discussed at length in \S\ref{s:abun_rats}, the
depletion and enhancement are linked through the adopted metallicity:
a lower metallicity implies a smaller C depletion and larger N
enhancement, and vice-versa.  I demonstrated in \S\ref{s:abun_rats}
and Figure~\ref{f:cn} that the derived C and N abundances for Algol
are consistent with the metallicity derived from O and Ne provided
that conservation of C and N nuclei held in the zone subjected to
CN-processing.  It therefore appears that the elements C, N, O and Ne
undergo very similar differential fractionation in both Algol and
HR~1099.

\subsection{Surface C and N abundance changes in HR~1099?}
\label{s:cnhr1099}

The XMM-Newton study of HR~1099 by Brinkman et al.\ (2001) and the
Chandra LETG study by Schmitt \& Ness (2002) both obtained C/N ratios
for HR~1099 that appear to be significantly lower than solar values
([C/N]$=-0.4$ relative to Anders \& Grevesse 1989; and [C/N]$< -0.16$
relative to ``cosmic abundances'', respectively).  Schmitt \& Ness
(2002) interpreted their result as indicating that the surface 
N abundance of HR~1099 has been modified by CN-cycle processing and
subsequent dredge-up (these authors noted an N abundance increase,
though the C/N ratio decreases primarily as a result of C depletion
because N would initially be $\sim 4$ times more abundant than C).
It would be interesting to see CN-cycle products in HR~1099 because the
X-ray dominant star is still a subgiant and has not yet crossed the
Herzsprung gap.  It should not, then, have a fully-developed deep
convective envelope and so dredge-up would have only just started.
Indeed, Lambert \& Ries (1981) noted that the subgiants in their
sample of evolved G and K stars appeared to have started the dredge-up
phase based on observed $^{12}C/^{13}C$ ratios.

If HR~1099 does indeed have modified surface C and N abundances, the
C/N ratio for Algol derived here relative to HR~1099 will need to be
modified: the change in C/N abundance ratio caused by CN-cycle
processing relative to its initial composition would be larger by the
magnitude of the depletion in the C/N ratio of HR~1099.  In this
regard, it is worth examining the Brinkman et al.\ (2001) and 
Schmitt \& Ness (2002) results in more detail.

The Brinkman et al.\ (2001) result is based on abundances derived
relative to O.  Since the temperature dependence of the emissivities
of the H and He-like lines of C, N and O are significantly different,
estimating the relative abundances of these elements based only on
these lines can be hazardous: errors in the adopted temperature
structure ($\Phi[N_e(T),V(T)]$) can induce errors in derived abundance
ratios.  Instead, the Schmitt \& Ness (2002) result is based on the
observed line strength ratio of the H-like Ly$_\alpha$ lines of C and
N, compared to the theoretical emissivity ratio as a function of
temperature.  Their upper limit to the C/N abundance ratio is based on
the maximum theoretical value for the line strength ratio (expressed
as N/C rather than C/N used here), and is therefore
temperature-independent.  Their maximum theoretical value is 0.42 in
photon units (0.57 in energy units), which is to be compared with
their measured ratio of $0.61\pm 0.05$.
The N/C line strength ratio computed with
the PINTofALE software (Kashyap \& Drake 2000), using the CHIANTI~v3
(Dere et al.\ 2001) implementation of the collision strengths of
Aggarwal \& Kingston (1991) together with the ionization balance of
Mazzotta et al.\ (1998) yields maximum values of the N/C line
strength ratio (in photon units) of 0.53, 0.59, 0.48 and 0.65
for the abundance mixtures of Allen (1973), Anders \& Grevesse (1989),
Grevesse \& Sauval (1998), and Grevesse \& Sauval (1998; N) combined
with Allende-Prieto et al.\ (2002; C), respectively.  My observed
line flux ratio, based on C and N line strengths from
Table~\ref{t:fluxes} normalised by the instrument effective areas at
the appropriate wavelengths (15.2 and 11.6~cm$^2$ for N and C, respectively;
Pease et al.\ 2002b) is $0.67\pm 0.07$.  This ratio is consistent with
that of Schmitt \& Ness (2002), but is also consistent with the   
theoretical upper limit for the N and C abundances of Grevesse
\& Sauval (1998) and Allende-Prieto et al.\ (2002), respectively, and
is only marginally higher than the upper limits corresponding to the
other abundance mixtures.  Based on this comparison, I then find no
firm evidence that the surface abundances of 
HR~1099 have been modified significantly by CN-processing.  The
difference between this conclusion and that of Schmitt \& Ness (2002)
lies predominantly in the adopted theoretical line ratio maximum and
in the assumed abundance mixture: for example, 
the C abundance of Allende-Prieto et al.\ (2002) is 0.17~dex lower
than that of Anders \& Grevesse (1989).

As pointed out by Schmitt \& Ness (2002; Figure 2) and Drake et al.\
(in preparation), the theoretical N/C line strength ratio is fairly flat for
temperatures $\log T\ga 6.4$.  For coronae dominated by emission from
plasma hotter than 2-$3\times 10^6$~K, such as that of HR~1099, one
can then interpret directly an observed N/C line strength ratio in
terms of the N/C abundance ratio.  The theoretical N/C ratio
corresponding to the Grevesse \& Sauval (1998) N abundance and
Allende-Prieto et al.\ (2002) C abundance for the temperature range $
6.4 \leq \log T \leq 7.2$ is $0.63\pm 0.03$.  Our observed ratio is
consistent with this value.  I therefore conclude that the ratio of C
and N abundances in the corona of HR~1099 is similar to that in the
Solar photosphere.  This also suggests that the inferred C/N ratio
from the abundances relative to O of Brinkman et al.\ (2001) might be
too low by a factor of $\sim 2$; again, 0.17~dex of this $\sim
0.3$~dex difference can be attributed to these authors having used the
Anders \& Grevesse (1989) abundance mixture.

Based on this result, and on the arguments cited above in
\S\ref{s:fractionate}, I conclude that there is unlikely to be any
differential compositional fractionation between C and N in the
coronae of HR~1099 or Algol.


%

\subsection{X-ray Emission from Algol A?}

In this analysis, I have assumed that there is not a significant
contribution to the LETGS spectrum from Algol A.  With a spectral type
of B8~V, Algol A is in the range of types from mid-B to late-A that do
not possess either strong radiatively-driven winds or outer convection
zones that are thought necessary to sustain shocks or magnetic
dissipation sufficient to heat and maintain plasma at X-ray emitting
temperatures.  These expectations were observationally confirmed for
B-type stars by Grillo et al. (1992) in an Einstein Observatory
survey.  In the few cases where X-ray emission was detected for late-B
and early-A stars, it was found to originate from a cool companion.
X-ray fluxes of up to $10^{31}$~erg~s$^{-1}$ were found for 86 dwarfs
of spectral types B7-B9 seen in the ROSAT All-sky Survey (Bergh\"ofer
et al. 1996), but it has since been discovered that these stars are
also either spectroscopic binaries or have young, lower mass companions
(e.g.\ Bergh\"ofer et al., 1997; Hubrig et al.\ 2001).  That late-B
and early-A main-sequence stars are X-ray dark is also supported by
the non-detection in X-rays of single A stars like Vega (Pease et al.\
in preparation; $< 9\, 10^{-4}$~count~s$^{-1}$ ROSAT PSPC; see also
Schmitt 1997).  In principle, then, X-ray wavelengths offer the chance
to obtain a clear view of the surface abundances of Algol secondary
stars.

While it then appears that {\em single} or {\em non-interacting}
late-B main-sequence stars are X-ray dark, it is plausible that the
accretion activity associated with Algol systems sustains hot plasma.
Singh et al.\ (1996) considered this problem in a comparison between
RS~CVn-type binaries comprised of two late-type stars in which
neither component filled their Roche lobes, and Algol-type binaries.
They found that the Algol-type binaries are in fact slightly X-ray
deficient relative to their RS~CVn cousins, suggesting strongly that
the accretion activity of Algols is not a significant source of
X-rays.  I discuss this further in \S\ref{s:comment}, below.

Evidence for plasma associated with accretion and with temperatures of
at least $10^5$~K has, however, been found in the Algol systems
V356~Sgr and TT~Hya from recent FUSE observations (Polidan et al.\
2000) and Peters et al.\ 2001, respectively).  Emission detected from
O~VI appeared to be associated with a bipolar flow that makes a large
angle with the orbital plane.  In order for this plasma to contribute
significantly at X-ray wavelengths, it must be comprised of components at
least an order of magnitude hotter than the formation temperature of
O~VI.  In the case of Algol, though the presence of such plasma cannot
be ruled out with certainty, Doppler shifts in X-ray lines also
suggest that Algol~B is the only significant source of X-rays.  A
recent investigation of spectral lines seen in the {\it Chandra} HETG
spectrum of Algol obtained in 2000 April clearly reveals the orbital
velocity of Algol~B and locates the X-ray emission on the late-type
star, leaving only a maximum of $\sim 10$\%\ of the observed line
emission as possibly arising from accretion in the vicinity of Algol A
(Chung et al., in preparation).

At this time, then, I conclude that significant X-ray emission from
sources other than the corona of Algol~B remains possible, though
unlikely, and that any such emission contributes at most 10\%\ of the 
total observed emission.  It also seems unlikely that such a
contribution would significantly change any of the conclusions derived
here, since the accreting material would be expected to have the same 
abundances of C and N as the corona of Algol~B.

\section{A Comment on the X-ray Luminosity Deficiency of Algol Binaries}
\label{s:comment}

In their comparative analysis of X-ray observations of Algol-type and
RS~CVn-type systems, Singh et al.\ (1996) noted that the Algols
appeared systematically fainter in X-rays than the RS~CVn's, even
after allowing for the likely X-ray dark nature of the primaries of
the former systems.  By comparing X-ray luminosity to Roche lobe
filling factors, Singh et al.\ (1996) ruled out mass transfer and
accretion in the Algol systems as significant factors in their X-ray
luminosity ``deficiency''.  Sarna et al.\ (1998) have successfully
explained the gross characteristics of the large X-ray luminosity
differences among Algol-type systems in terms of convection zone
parameters responsible for generating magnetic flux within an
$\alpha$-$\omega$ dynamo formalism.  Nevertheless, it is not clear
that the systematic trend of the RS~CVn's appearing systematically
more X-ray bright than the Algols, as illustrated in Figures~4a and 4b
of Singh et al.\ (1996), can be entirely explained within such a
framework.

For what other reasons might what are otherwise very similar
star systems have notably different X-ray characteristics?  One
conspicuous difference between the RS~CVn's and the Algols is the mass
transfer history of the secondaries of the latter.  Loss of a
substantial fraction of the outer envelope leads to less dilution of
the layers processed by main-sequence nuclear burning, as is manifest
in the C/N abundance ratio.  While trace elements such as C and N will
have little effect on a star or its corona, another element whose
surface abundance can be substantially altered is He.  

It was shown by Drake (1998) that a corona substantially enhanced in
He might appear metal-poor owing to the dependence of thermal
bremsstrahlung on the square of the nuclear charge: the enhanced
continuum lowers the line-to-continuum ratio.  In the Algol case,
however, Drake (1998) also noted that 
the large He abundance should not appreciably change the
emitting characteristics of the corona.  This is because the
enhancement in He occurs at the expense of the loss of H, where four H
nuclei are transformed into one He nucleus.  In this case, the
line-to-continuum ratio remains about the same because the He
continuum emissivity is four times that of H.  The net emissivity of
coronal plasma therefore remains about the same.

Instead, I speculate that raising the He abundance could affect the
corona in two possible ways: (i) through changes in the convective
envelope in which magnetic fields thought to fuel a corona are
generated; (ii) through a possible dependence of the coronal heating
efficiency on the He abundance or plasma mean molecular weight.
Suggestion (i) would have been implicitly included in the work of
Sarna et al.\ (1998), and a comparison between convection zone
parameters of otherwise very similar systems differing only in the He
abundance in the envelope might prove interesting.  In the case of the
latter scenario, that the He abundance in the corona might directly
affect the coronal heating efficiency, Algol systems might prove
valuable for helping discriminate between different coronal heating
mechanisms.

\section{Summary}
\label{s:summary}

I have investigated the C and N abundances in the corona of Algol~B
through a differential study relative to HR~1099.  Chandra LETG
spectra obtained for both stars during times of quiescence 
show that the temperature structure and abundances in these
coronae are very similar, in accordance with naive expectations based
on the similarity of the parameters of the stars that dominate the
X-ray emission.  The differential analysis presented here therefore
avoids errors associated with determining the underlying DEM, in
addition to all systematic uncertainties involving atomic data, line
flux measurement and instrument calibration.

The C/N abundance ratio in HR~1099 is found to be consistent with that
of the solar photosphere, and no evidence is found for modification of
the surface abundances of the X-ray-dominant subgiant through CN-cycle
processing.  The difference between this conclusion and that of
Schmitt \& Ness (2002), who concluded that these elements exhibited
signs of CN-processing, can be attributable to adoption of a different
solar abundance for C.  In contrast, C is not detected in the corona
of Algol and N appears enhanced relative to HR~1099 by a factor of 2.
The change in the C/N abundance ratio in the atmosphere of Algol~B as
a result of CN-cycle processing is found to be $\Delta\log C/N \leq
-1.44$.  Based on this result, a simple two-zone model demonstrates
that Algol~B must have lost at least half of its original mass.  This
result is also consistent with predictions of detailed evolutionary
models for Algol computed by Sarna (1993) in which Algol~B has an
initial mass of $3.5M_\odot$ and loses about $2.7M_\odot$.

A detailed examination of the abundance patterns in HR~1099 and
Algol~B for elements Fe, Ne and O confirms that they have very similar
fractionation patters, consistent with the view that they are
``coronal twins''.  Consideration of the likely photospheric
abundances for these stars leads to the conclusion that Fe must be
depleted in their coronae relative to photospheric values by about
0.5~dex.  One notable difference in their coronal properties
will be an enhanced He abundance in the corona Algol~B owing
to the mixing of the products of core H-burning into the remnant
envelope.  I speculate that the enhanced He abundance in some Algol
systems might explain some of the previously observed X-ray deficiency
of these systems as compared to similar RS~CVn-type binaries.

\acknowledgments

I would like to express warm gratitude to Dr.~V.~Kashyap for helpful
comments, and to M.~Sarna for communicating his unpublished model N
abundances for Algol~B and for his stimulating remarks.  A helpful
review from an anonymous referee enabled me to improve the manuscript
and include some relevant issues I had not considered.  I also thank
the NASA AISRP for providing financial assistance for the development
of the PINTofALE package, and the CHIANTI project for making publicly
available the results of their substantial effort in assembling atomic
data useful for coronal plasma analysis.  JJD was supported by NASA
contract NAS8-39073 to the {\em Chandra X-ray Center} during the
course of this research.

\clearpage


\newpage\section*{Tables}

\begin{deluxetable}{lrcccccrrr}
\tabletypesize{\scriptsize}
\tablecaption{Summary of Stellar Parameters and 
Chandra LETG+HRC-S Observations\label{t:obs}}
\tablewidth{0pt}
\tablehead{ \colhead{} & \colhead{Spectral} & 
\colhead{Period} & \colhead{Distance} 
& \multicolumn{3}{c}{Active Component Parameters\tablenotemark{a}}
& \colhead{Obs} & \colhead{Start} & \colhead{Exposure}\\
\colhead{Star} & \colhead{Type} & \colhead{[d]} & \colhead{[pc]} 
& \colhead{T$_{eff}$[K]}& \colhead{$R_\odot$} & \colhead{$M_\odot$}
&\colhead{ID}& \colhead{[UT]} & \colhead{[s]}
} 
\startdata
Algol 	& B8 V + G8 III &  2.87\tablenotemark{b} & 29.0\tablenotemark{d} & $4500\pm 300$\tablenotemark{efg} &3.5\tablenotemark{eg} & 0.8\tablenotemark{eg} & 0002 & 2000-03-12T18:23:15 &
79531 \\
HR~1099 & K1 IV + G5 V  &  2.84\tablenotemark{c} & 28.5\tablenotemark{d} & $4750\pm 200$\tablenotemark{h} &3.7\tablenotemark{i} &1.0\tablenotemark{i} & 1879 & 2001-01-10T23:14:49 &
94867 \\
\enddata
\tablenotetext{a}{The component that dominates X-ray Activity: Algol B
and the K1~IV component of HR~1099.}
\tablenotetext{b}{Soderhjelm (1980)}
\tablenotetext{c}{Fekel (1983)}
\tablenotetext{d}{Perryman et al.\ (1997)}
\tablenotetext{e}{Richards et al.\ (1988)}
\tablenotetext{f}{Eaton (1975)}
\tablenotetext{g}{Kim (1989).  
This work obtained $4888\pm 96$~K; type K0~IV was
assigned to the secondary based on this temperature, combined with the
values of the mass and radius similar to those of Richards et al.\
(1988).}
\tablenotetext{h}{Randich et al.\ (1984)}
\tablenotetext{i}{Donati (1999), who notes that these values are 
for $i=38\degr$ and are somewhat sensitive to the
inclination adopted; in particular the mass and radius are larger for
smaller inclinations.}
\end{deluxetable}


\begin{deluxetable}{lllllllll}
\tabletypesize{\scriptsize}
\tablecaption{Identifications and Fluxes for Spectral Features Used in
this Analysis\label{t:fluxes}} 
\tablehead{
\colhead{$\lambda_{\rm obs}$} & 
\colhead{$\lambda_{\rm pred}$} &
\colhead{Ion} & 
\colhead{$\log T_{\rm max}$\tablenotemark{a}}
 & \colhead{Algol} 
 & \colhead{HR1099} 
 & \colhead{Ratio\tablenotemark{b}} 
 & \colhead{Transition}
 & \colhead{Notes}
 \\
\colhead{(\AA )} &
\colhead{(\AA )} &
\colhead{} &
\colhead{(K)}
 & \colhead{[Counts]} 
 & \colhead{[Counts]} 
 & \colhead{} 
 & \colhead{} 
 & \colhead{} 
}

\startdata

15.019 & 15.015 & Fe XVII& 6.75 & $1179 \pm 54$ & $1315 \pm 54$ &
$0.89\pm 0.06$ & 
(2p$^5$.3d) $^1$P$_1$ $\rightarrow$ (2p$^6$) $^1$S$_0$ \\
18.957 & 18.967 & O VIII & 6.50 & $2877 \pm 98$ & $5612 \pm 141$ & $0.513 \pm 0.022$ & (2p) $^2$P$_{3/2},_{1/2} \rightarrow$ (1s) $^2$S$_{1/2}$ & \\
20.914 & 20.910 & N VII	 & 6.35 & $ 148 \pm 49$ & $88   \pm 30$ & $1.69 \pm 0.80$ & (3p) $^2$P$_{3/2}$ $\rightarrow$ (1s) $^2$S$_{1/2}$ & \\
21.608 & 21.602 & O VII	 & 6.30 & $ 290 \pm 35$ & $557  \pm 53$ & $0.522 \pm 0.080$ & (1s.2p) $^1$P$_1$ $\rightarrow$ (1s$^2$) $^1$S$_0$ & \\
24.252 & 12.132 & Ne X	 & 6.80 & $ 133 \pm 28$ & $244  \pm 45$ & $0.55 \pm 0.15$ & (2p) $^2$P$_{3/2}$ $\rightarrow$ (1s) $^2$S$_{1/2}$ & 2nd order \\
24.785 & 24.779 & N VII	 & 6.30 & $1177 \pm 56$ & $632  \pm 53$ & $1.85 \pm 0.18$ & (2p) $^2$P$_{3/2}$ $\rightarrow$ (1s) $^2$S$_{1/2}$ & \\
33.744 & 33.734 & C VI	 & 6.20 & $  < 44 $     & $718  \pm 53$ & $<0.064 $ & (2p) $^2$P$_{3/2}$ $\rightarrow$ (1s) $^2$S$_{1/2}$ & \\
36.398 & 12.132 & Ne X	 & 6.80 & $ 273 \pm 32$ & $430  \pm 41$ & $0.63 \pm 0.10$ & (2p) $^2$P$_{3/2}$ $\rightarrow$ (1s) $^2$S$_{1/2}$ & 3rd order \\
37.920 & 18.967 & O VIII & 6.50 & $ 130 \pm 27$ & $242  \pm 43$ & $0.54 \pm 0.15$ & (2p) $^2$P$_{3/2}$ $\rightarrow$ (1s) $^2$S$_{1/2}$ & 2nd order \\
26.75 & \nodata & Cont & \nodata & $7530 \pm 87$	& $9065 \pm 95$ &
$0.83 \pm 0.01$ & b-f, f-f; 25.5-28.0 \AA & Continuum \\
\enddata
\tablenotetext{a}{The temperature at which the function $G_{ij}(T)$
peaks (Eqn.~\ref{e:flux})}
\tablenotetext{b}{The ratio of observed integrated counts; this is 
proportional to the ratio of fluxes $F_{A}/F_{H}$ 
(Eqn.~\ref{e:flux_rat}).}
\end{deluxetable}


\begin{deluxetable}{lllllllll}
\tablecaption{Derived Abundances Relative to 
HR~1099\tablenotemark{a}\label{t:abuns}} 
\tablehead{
   \colhead{[Fe/H]$_{1099}$}
 & \colhead{[O/H]$_{1099}$}
 & \colhead{[Ne/H]$_{1099}$}
 & \colhead{[M/H]$_{1099}$\tablenotemark{b}}
 & \colhead{[C/M]$_{1099}$}
 & \colhead{[N/M]$_{1099}$}
}
\startdata
$0.03 \pm 0.03$ & $-0.21 \pm 0.02$ & $-0.13 \pm 0.06$ & $-0.20\pm 0.02$ & 
$< -0.93$\tablenotemark{c} & $0.51 \pm 0.02$ \\
\enddata
\tablenotetext{a}{Expressed in the conventional logarithmic bracket
notation where [X/H] represents the abundance of element X relative to
a standard star (usually the Sun, though here relative to HR~1099).
{\bf NB: Quoted uncertainties are based on propagation of statistical
uncertainties only; true errors are expected to be of order 0.1~dex.}}
\tablenotetext{b}{The error-weighted mean of the O and Ne abundances
relative to HR~1099.}
\tablenotetext{b}{$3\sigma$ upper limit.}
\end{deluxetable}

\newpage

\end{document}